\documentclass[12pt,thmsa]{article}
\usepackage{sw20aip}

\input{tcilatex}

\begin{document}

\author{O. Sotolongo-Costa$^{1}$, F. Guzman$^{2}$, J. C. Antoranz$^{3}$,  \and G. J.
Rodgers$^{4}$, O. Rodriguez$^{2}$, J.D.T. Arruda Neto$^{5}$, \and A. Deepman$%
^{5}$ \and {\small 1.- Department of Theoretical Physics. U.H., Havana 10400,%
} \and {\small Cuba. \ \TEXTsymbol{<}oscarso@ff.oc.uh.cu\TEXTsymbol{>}} \and 
{\small 2.- Institut of Nuclear Sciences and Technology, Havana 10600,} \and 
{\small Cuba.\TEXTsymbol{<}guzman@info.isctn.edu.cu\TEXTsymbol{>}} \and 
{\small 3.- Departamento de Fisica Matematica y Fluidos, UNED, 28040} \and 
{\small Madrid. \TEXTsymbol{<}antoranz@apphys.uned.es\TEXTsymbol{>} } \and 
{\small 4.-Department of Mathematical Sciences. Brunel University,} \and 
{\small Uxbridge, UK.\TEXTsymbol{<}g.j.rodgers@brunel.ac.uk\TEXTsymbol{>}}
\and {\small 5.- Instituto de F\'{i}sica, Universidade de Sao Paulo, Brasil,
CEP:} \and {\small 05315-970 \TEXTsymbol{<}arruda@usp.br\TEXTsymbol{>}}}
\title{A non extensive approach for DNA\ breaking by ionizing radiation.}
\maketitle

\begin{abstract}
Tsallis entropy and a maximum entropy principle allows to reproduce
experimental data of DNA double strand breaking by electron and neutron
radiation. Analytic results for the probability of finding a DNA\ segment of
length $l$ are obtained reproducing quite well the fragment distribution
function experimentally obtained.
\end{abstract}

Atomic force microscopy (AFM) has revealed itself as an extremely useful
device in the analysis of very small structures and specially in DNA
fragment analysis as it has been shown in \cite{berman}.

It is interesting to study the production of fragments in DNA as a result of
radiation, since the presence of radiation interacting with DNA\ molecules
can influence the properties of living cells up to a lethal extreme.

On the other hand, DNA\ fragment analysis may help in the study of the
structural properties of genome texts, and then to understand general
principles of genetic sequences.

The process of DNA double strand breaking was performed by irradiation of
DNA\ molecules with electrons and neutrons at different doses ( See \cite
{berman} ). Then, the length of the resulting fragments was measured. As a
result, the collection of fragments was found to obey a fragment size
distribution function (FSDF), which presents important characteristics from
the viewpoint of complexity.

The main fact, which will be focused in this paper, is that the collection
of fragments is such that there is not a ``characteristic'' size of the
fragments, \textit{i.e}., the smaller the fragment, the more abundant is it.
The FSDF in this case does not present a definite local maximum, resembling
more to an inverse power law, i.e., a distribution function in the basin of
attraction of a stable (L\'{e}vy) distribution \cite{feller}.

The main distinction of L\'{e}vy distributions lies in the fact that their
variance is divergent. Maybe because of it, scientists have paid attention
to them only recently.

This feature of the FSDF is not new. It has been reported in \cite{ishii}
the occurrence of transition to scaling in FSDF during glass rods breaking,
in \cite{oddershede} the power law distribution of fragments was related to
self-organized criticality (SOC). Matsushita \cite{matsu} proposed a fractal
representation for a general process of fragmentation.

Our group \cite{pre,prl} detected power law behavior in the process of
liquid drop fragmentation and we proposed a Bethe lattice representation to
interpret FSDF in these experiments.

Some attempts to relate FSDF to first principles in physics like the maximum
entropy principle are present in \cite{englman,li-tankin} with results that,
at the best, do not cover the process in which scaling in FSDF\ is present. (%
\textit{i.e.}, when the energy of the fragmentation process is high).

The universal nature and almost unlimited range of applicability of the
maximum entropy principle leads us to expect it to be useful in describing
scaling in FSDF even at DNA\ scale.

But the process of fractionating, by its own nature, is a paradigm of
phenomena in which interactions are long-range correlated among all parts of
the object under fragmentation. Then, though the maximum entropy principle
is expected to have an unlimited range of application, in the process of
breaking the expression for the entropy in its Shannon form:

\begin{equation}
S=-k\int p(x)\log p(x)dx  \label{eq:1}
\end{equation}
-where $p(x)dx\label{eq:1}$is the probability of finding the system
magnitude $x$ in the interval $[x,x+dx]$ , and $k$ is Boltzmann's constant-
is not applicable.

This is because this formula, based in Boltzmann-Gibbs statistics, is
expected to be valid when the effective microscopic interactions are
short-ranged, and this gives to this entropy its extensive character (The
entropy of the whole object equals the sum of the entropies of its
constituent independent parts).

Since, as we already pointed out, all parts of the fractionating object
during the process of violent breakage are correlated, then the entropy of
the object being fractionated is smaller tan the sum of the entropies of the
parts in which the object divides, defining this way a ``superextensivity''
in this system. This suggests that it may be necessary to use non-extensive
statistics, instead of the Boltzmann-Gibbs one.

This kind of theory has already been proposed by Tsallis \cite{tsallisjsp},
who postulated a generalized form of entropy, given by

\begin{equation}
S_q=k\frac{1-\int p^q(x)dx}{q-1}  \label{eq:2}
\end{equation}

where q is a real number.

This entropy can also be expressed as:

\begin{equation}
S_q=-k\int p(x)l_qp(x)  \label{eq:3}
\end{equation}

where the generalized logarithm $l_qp(x)$ is defined as (See \cite
{tsallisbjp}):

\begin{equation}
l_q(p)=\frac{p^{1-q}-1}{1-q}  \label{eq:4}
\end{equation}

It is straightforward to see that $S_q\longrightarrow S$ when $%
q\longrightarrow 1$ , recovering Boltzmann-Gibbs statistics.

It is our goal to derive, starting from first principles, a functional
dependence to describe the DNA DSDF obtained in \cite{berman}.

Starting from equation \ref{eq:2} we may follow the method of Lagrange
multipliers to apply the maximum entropy principle to the fragmentation of
DNA. To do this, we impose two constraints: The first is the trivial one of
normalization of the probability:

\begin{equation}
\int p(l)dl=1  \label{eq:5}
\end{equation}

\textit{i.e., }the sum of the probabilities of finding a fragment of any
length is equal to unity.

As a second constraint we may choose to adopt a ``q-mean value'' as:

\begin{equation}
\int p^q(l)ldl=1  \label{eq:6}
\end{equation}

Which reduces to the classical mean value when $q\longrightarrow 1$ .In this
formulation the length $l$ of the fragments has been referred to a unit
adequately chosen as to choose the ''q-mean value'' equal to one.

It may seem strange to introduce a ``q-mean value'' , also known as
``unnormalized mean value'' in this formulation. Really, this choice is not
unique but for our purposes and for simplicity reasons we will choose this
formulation. In \cite{tsallisbjp} a detailed discussion of the possible
choices for the second constraint can be found. The one here chosen showed
to be particularly useful in describing anomalous diffusion and was also
employed by us in \cite{entropy,physA} dealing with problems of
fragmentation.

Now we use the method of Lagrange multipliers by means of the construction
of the functional:

\begin{equation}
\pounds (p_i,\alpha ,\beta )=S_q-\alpha \int p(l)dl+\beta \int p^q(l)ldl
\label{eq:7}
\end{equation}

being $\alpha $ and $\beta $ the Lagrange multipliers.

The extremization of this functional leads to:

\begin{equation}
p(l_i)=\frac{\beta (2-q)dl}{[1+\beta (q-1)l]^{1/(q-1)}};  \label{eq:8}
\end{equation}

Alternatively, the same method when applied to the Boltzmann entropy ($q=1$)
gives

\begin{equation}
p(l)dl=\beta e^{-\beta l}dl.  \label{eq.9}
\end{equation}
\label{eq.9}Equation \ref{eq:8} is the expression for the probability of
finding a fragment of length $l_i$ and depends on three coefficients to
adjust. In this case we can apply this expression to fit it to the
experimental data reported in \cite{berman}., where methods of atomic force
microscopy were applied to measure FSDF of irradiated DNA.

Figure 1 shows the experimental results for DNA breaking with electrons at
doses of 5000 and 7000 Gy. Both are fitted with Eq. \ref{eq:8}. Figure 2
represents FSDF for DNA breaking with neutrons at the same doses. In both
cases the length of the fragments was normalized to the length of the
largest one, and the number of fragments was normalized to the total number
of fragments. As it can be seen, the agreement is very good.

More experimental data for electrons from 50 to 200 Gy and neutrons at doses
of 900, 7500, 2000 and 10000 Gy were also fitted with good results. In this
paper we are reporting the results of the coincident doses of electrons and
neutrons of 5000 and 7000 Gy to illustrate the application of this
viewpoint. Only in the cases of very low doses of electrons (50 and 100 Gy,
where fluctuations in FSDF are important) the results are not as good as the
ones before.

This fact reveals the non extensive nature of DNA breaking, as it was shown
for macroscopic objects in \cite{entropy,physA}. So, This characteristic of
breaking is not exclusive of macroscopic bodies.

Use of Boltzmann\'{}s entropy to describe FSDF obtained in these experiments
leads to incorrect results, (\textit{i.e}.,\ref{eq.9}) impossible to fit
with the data, which shows power law behavior.

On the other hand, this non extensivity may also reflect an intrinsic nature
of the very DNA chain. The presence of $1/f$ spectrum in sequences and
long-range correlations in the DNA\ sequences \cite{voos} supports this
assertion.

This work has been partially supported by the ``Alma Mater'' contest, Havana
University. One of us (O.S.) is indebted to the Department of Mathematical
Physics and Fluids, UNED, for kind hospitality.

\subsection{Figure Captions}

Fig.1: Normalized DSDF for electron irradiation of DNA at 5000 Gy and 7000
Gy. The solid squares represent the experimental results at 7000 Gy and
solid circles at 5000 Gy. Fitting was made with Equation \ref{eq:8}. Solid
curve is for 5000 Gy. Curve with open squares is for 7000 Gy.

Fig. 2: Normalized DSDF for neutron irradiation of DNA at 5000 Gy and 7000
Gy. The solid squares represent the experimental results at 7000 Gy and
solid circles at 5000 Gy. Fitting was made with Equation \ref{eq:8}. Solid
curve is for 5000 Gy. Curve with open squares is for 7000 Gy.

\end{document}